\documentclass[prb,twocolumn]{revtex4-1} 
\usepackage{amsmath}  % needed for \tfrac, \bmatrix, etc.
\usepackage{amsfonts} % needed for bold Greek, Fraktur, and blackboard bold
\usepackage[dvipdfmx]{graphicx}

\newcommand*\encircle[1]{\unitlength1ex\ \begin{picture}(2.5,2.5)%
\put(0.75,0.75){\circle{2.7}}\put(0.75,0.75){\makebox(0,0){#1}}\end{picture}}

\newcommand*\ensquare[1]{\unitlength1ex\ \begin{picture}(2.5,2.5)%
\put(-0.5,-0.5){\framebox(2.4,2.4)}\put(0.75,0.75){\makebox(0,0){#1}}\end{picture}}

\begin{document}

% Be sure to use the \title, \author, \affiliation, and \abstract macros
% to format your title page.  Don't use lower-level macros to  manually
% adjust the fonts and centering.

\title{Exploring cogging free magnetic gears}
% In a long title you can use \\ to force a line break at a certain location.

\author{Stefan Borgers}
\author{Simeon V\"olkel}
\author{Wolfgang Sch\"opf}
\email{wolfgang.schoepf@uni-bayreuth.de} % optional
\author{Ingo Rehberg}
\email{ingo.rehberg@uni-bayreuth.de} % optional
\affiliation{Experimental Physics V, University of Bayreuth, 95440 Bayreuth, Germany}

% See the REVTeX documentation for more examples of author and affiliation lists.

\date{\today}

\begin{abstract}
The coupling of two rotating spherical magnets is investigated experimentally, with particular 
emphasis on those motions where the driven magnet follows the driving one with a uniform angular 
speed, which is a feature of the so called cogging free couplings. The experiment makes use of 
standard equipment and digital image processing. The theory for these couplings is based on 
fundamental dipole-dipole interactions with analytically accessible solutions. Technical 
applications of this kind of coupling are foreseeable particularly for small machines, an 
advantage which also comes handy for classroom demonstrations of this feature of the fundamental 
concept of dipole-dipole coupling.
\end{abstract}
% AJP requires an abstract for all regular article submissions.
% Abstracts are optional for submissions to the "Notes and Discussions" section.

\maketitle % title page is now complete

\section{Introduction} 
\label{intro}
Teaching physical interactions normally starts with Newton's gravitation law, which paves the way
towards Coulomb's law for the electromagnetic interaction. While in principle this second interaction
is the dominant one on the scale of a classroom, its strength is suppressed by the fact that matter
generally does not carry a net charge. This force rather appears as the net force of a distribution of
charges, which can be described in leading order as the interaction of dipoles. Thus the relevance of
dipole-dipole interaction for the description of the interaction of matter can hardly be 
overestimated.\cite{hirschfelder1964}
This justifies its fundamental role in science education, which is facilitated
by the fact that experimental investigations can be done with a small budget due to the availability of
strong spherical magnets.\cite{adams2007} Those magnets are fortunately equivalent to electric 
dipoles,\cite{griffiths1992} so that the underlying mathematics based on the interaction potential of point
dipoles is elementary accessible.\cite{edwards2017}

Already the static interaction of such dipoles provides fundamental problems.\cite{luttinger1946,belobrov1983} 
For a small number of spherical magnets, some of these questions can be tackled
experimentally within classroom demonstrations. Can I build an arrangement of spheres reminiscent
of a monopole?\cite{vandewalle2014} How many stable arrangements does the smallest 
three-dimensional cluster, the tetrahedron, have,\cite{schonke2015a} and what is the energetically
preferred orientation of a small number of dipoles anyway?\cite{messina2014} 
How many spheres can I stack on top of each other before gravity destroys this
tower?\cite{vella2014,schonke2017}

The number of phenomena is naturally even richer when it comes to the dynamics of interacting
dipoles. The chaotic pendulum might be the most prominent example in this field.\cite{sprott2006}
The coupling of only two dipoles provides a nonlinear oscillator with a beautiful analytic solution for
its periodic oscillations.\cite{pollack1997} Making use of the potential energy stored in a clever
arrangement of magnetic dipoles allows the construction of the so-called Gauss rifle.\cite{chemin2017}
The fact that students are fascinated by this device manifested recently in a happening presenting a
$546\,$m long rifle of this kind, which achieved a ``Guinness World Record''.\cite{unimainz2017}

Finally, technical applications\cite{nagrial1993} might also be a stimulating motivation to learn about
dipole-dipole interaction, a route which is stressed in the current paper. In fact, the milk frother you
might have used this morning to create the topping for your coffee might contain a magnetic gear
similar to the ones used in chemistry labs in the form of a magnetic stirrer. Both devices make use of
the fact that magnetic gears are free of contact, so that the input and output can be physically
isolated. Other advantages of such gears in comparison to mechanical ones are that they are not
subject to mechanical wear, need no lubrication, possess inherent overload protection, are noiseless
and reliable.

\begin{figure}[b]
\centering
\includegraphics[width=0.8\columnwidth]{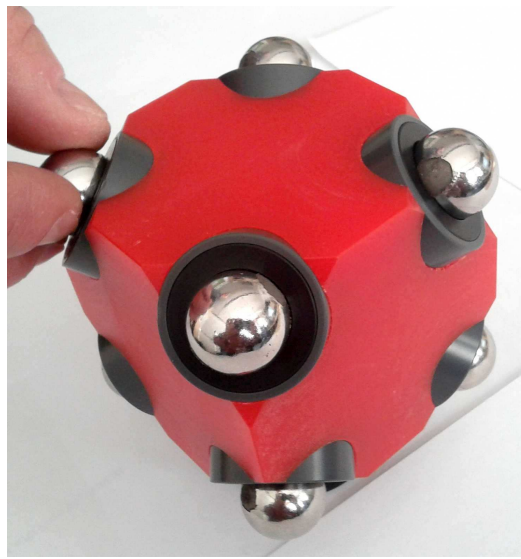}
\caption{The ``magnetic cube''. Eight spherical magnets are located at the corners of a
cube and can rotate freely in their positions fixed by the cube.}
\label{fig:cube}
\end{figure}

A particularly simple design for a classroom demonstration of a magnetic coupling is presented in 
Fig.~\ref{fig:cube}.\cite{schonke2015a} It contains eight spherical magnets located at the corners 
of a cube, which is done here by providing eight holes in the corners of a Makrolon\textregistered\  
cube. Because the spheres easily rotate inside the holes (additional ball bearings are helpful but 
not crucial), they will arrange automatically and find their position in the magnetic ground state, 
where they attract each other. When turning one of the spheres around the diagonal axis of the cube, 
the other seven rotate as well, all around their freely chosen axes, namely their corresponding space 
diagonal. The rotation can be performed without the need to overcome a magnetically induced mechanical 
resistance. This technical advantage is known as \textit{cogging free} in the context of magnetic 
motors and gears. The physical reason lies in the fact that the magnetic ground state of this 
eight-dipole arrangement forms a degenerate continuum with respect to dipole rotations around the 
space diagonal. Although this is a rather nontrivial demonstration of a cogging free magnetic clutch 
from a mathematical point of view,\cite{schonke2015a} it has the charm that this simple apparatus can 
be assembled directly in the classroom (by preferably four students working together, because eight 
hands are needed during the assembly process). The machinery described in the present paper contains 
only two magnetic spheres. It is conceptually simpler and presumably more suitable for technical 
applications than the sevenfold magnetic clutch shown in Fig.~\ref{fig:cube}, but it is mechanically 
costlier due to the need of additional fixed rotation axes.

The theoretical description of such gears designed with spherical magnets can be based on 
the concept of pure dipole-dipole interaction for two dipoles $\vec{m}_1$ and $\vec{m}_2$. 
The coupling of two rotating dipoles with predefined rotation axis has been discussed 
by Sch\"onke.\cite{schonke2015b} He proposed a family of geometrical arrangements of the two 
rotation axes, for which the coupling should be cogging free. One type of coupling, the so-called trivial one, is
reminiscent of the well-known coupling in magnetic stirrers, while the nontrivial one is characterized
by the fact that the driven sphere changes its sense of rotation when compared to the trivial one.
The latter one did not yet lead to real technical application, but it seems worth mentioning that a
school student recently participated in a scientific competition with a toy car making use of this 
drive.\cite{jugendforscht2017} In the current paper the first measurement of both couplings within a single
experimental setup is presented, together with a theoretical description based on dipole-dipole
interaction.

\section{Experimental Set-up}
\label{setup}
\begin{figure}[b]
\centering
\includegraphics[width=\columnwidth]{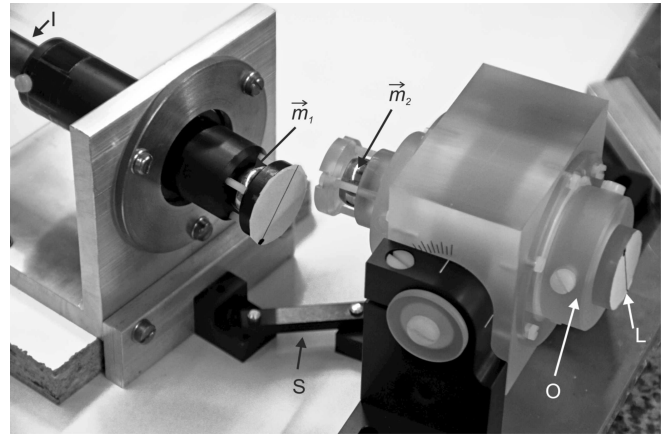}
\fbox{\includegraphics[width=0.975\columnwidth]{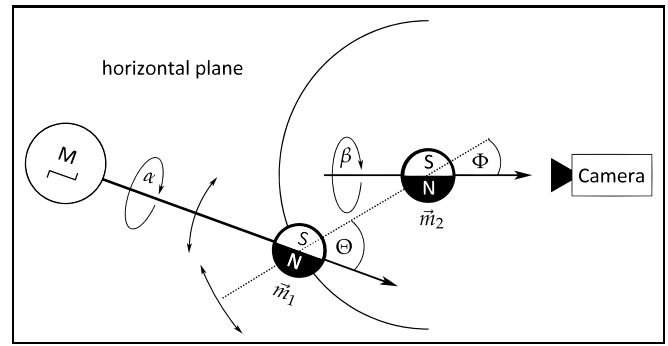}}
\caption{\textbf{Top:} Photograph of the experimental setup with the input shaft I, the output shaft O, the two
dipoles $\vec{m}_1$ and $\vec{m}_2$, the spacer S and the line L (for further explanations, see main text). Not shown
here are the motor that turns the input shaft and the camera which records the line.\\
\textbf{Bottom:} Top view schematic of the experimental setup with all parts remaining in the same horizontal plane.
A computer controlled stepper motor M rotates the input shaft with the dipole $\vec{m}_1$ by the 
\textit{input dipole angle} $\alpha$, thus giving rise to a rotation of the output shaft with the dipole $\vec{m}_2$ 
by the \textit{output dipole angle} $\beta$. $\beta$ is detected by a CCD-camera which is also controlled by the PC. 
The \textit{input shaft angle} $\Theta$ denotes the angle between the input shaft and the dotted line connecting the two 
dipoles, while the \textit{output shaft angle} $\Phi$ is the respective angle for the output shaft. The center of 
$\vec{m}_2$ is fixed in space and the distance between $\vec{m}_1$ and $\vec{m}_2$ is kept constant by a spacer. 
The center of $\vec{m}_1$ together with the input shaft can be shifted on a circular arc around the center of 
$\vec{m}_2$ such, that $\Theta$ is kept constant but $\Phi$ is changed. The input shaft can be turned around 
the center of $\vec{m}_1$, thus changing $\Theta$.}
\label{fig:setup}
\end{figure}

A photograph of the setup is given in Fig.~\ref{fig:setup}, top, while a schematic top view of the 
horizontal plane in which the experiments take place is shown in Fig.~\ref{fig:setup}, bottom. 
We study the interaction of two rotating 
dipoles in the following geometry:\cite{volkel2016,borgers2016} each dipole is attached to a shaft, 
with the dipole moments being perpendicular to the axis of the respective shaft, and can rotate about 
the respective axis. Both axes always remain in the same horizontal plane. One of the shafts 
(the \textit{drive} or \textit{input unit}) is actively driven by a stepper motor,
while the other shaft (the \textit{output unit}) can rotate freely. The two dipoles are realized 
by spherical neodymium permanent magnets with diameters of $19\,$mm each. 
They are attached to the shafts such, that their dipole moments are perpendicular to the respective axis
with a precision of about $3^\circ$. 
The comparison of the magnetic field of such a sphere with that of an ideal point dipole is given in
appendix \ref{app:dipole}.

Due to the strong magnets and the forces they exert on the system, the shaft bearings need special 
attention. We use non-magnetic and electrically non-conducting bearings. The input shaft 
runs in a bearing consisting of a plastic cage with glass beads, which was constructed in our
in-house workshop, and is connected to the stepper motor by an $80\,$cm long brass rod. In this way,
the motor and the magnets do not interfer magnetically. The output shaft runs in an industrial
ceramic bearing made from zirconium oxide. In order to visualize the orientation of the dipole $\vec{m}_2$,
a marker in form of a black line on a white background is attached to the end of the output shaft. 
This line is recorded by a CCD-camera.

The stepper motor and the CCD-camera are both connected to a computer, so that the input dipole angle
$\alpha$ (through the stepper motor) and the output dipole angle $\beta$ (by digital image processing)
are recorded and can be evaluated. The output shaft is fixed to a granite table, while the input
shaft can be freely oriented on this table. A spacer is used to keep the distance between the dipoles 
constant during a set of experiments, while the relative orientation of the two axes can be varied.

\begin{figure}[t]
\centering
\includegraphics[width=1.0\columnwidth]{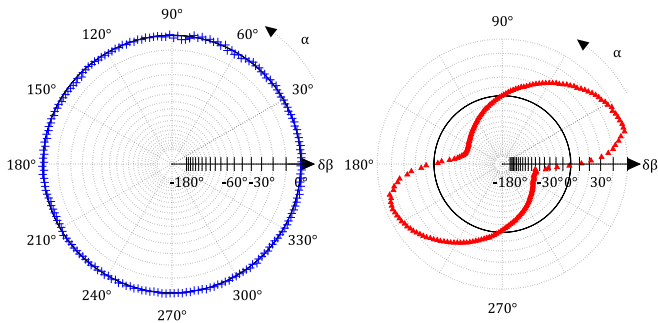}
\caption{Phase jitter $\delta \beta$ as a function of the input dipole angle $\alpha$ for a cogging free
(left; $\Theta \approx \Phi \approx 0^\circ$) and for a non cogging free coupling (right; $\Theta \approx -16^\circ$ 
and $\Phi \approx -61^\circ$).}
\label{fig:motivation}
\end{figure}

Figure~\ref{fig:motivation} shows two examples of quasi-static measurements in polar coordinates. The angular
coordinate represents the input dipole angle $\alpha$, while the radial coordinate shows the phase jitter
$\delta \beta$, which is the deviation of $\beta$ from the value expected for a perfectly phase-locked 
operation and characterizes the deviation from a cogging free motion as described in detail in the 
following section. The left image corresponds to $\Theta \approx \Phi \approx 0^\circ$, an arrangement
which is realized in magnetic stirrers with no cogging between the two magnets. The circle indicates, 
that $\beta$ follows $\alpha$ almost perfectly, but the phase jitter plot cannot show a constant phase 
shift, which in this case is $180^\circ$. The general case, however, is different with an example shown 
in the right image for $\Theta \approx -16^\circ$ and $\Phi \approx -61^\circ$. 
Here, $\beta$ depends in a nonlinear fashion on $\alpha$. Such a motion is accompanied by non-zero cogging.

\section{Theoretical Background}
\label{theory}
\begin{figure}[b]
\centering
\includegraphics[width=0.8\columnwidth]{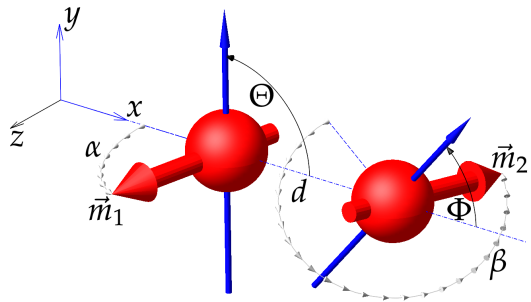}
\caption{The coordinate system with the relevant angles. 
The $x$-axis points from the center of the input dipole (left red sphere) to the center of the output dipole
(right red sphere), which are a distance $d$ apart. Both rotation axes (blue arrows) lie in the $x$-$y$-plane 
and enclose with the $x$-axis the angles $\Theta$ and $\Phi$, respectively. $\Theta$ is defined to range from 
$-180^\circ$ (negative $y$) to $+180^\circ$ (positive $y$), while $\Phi$ only ranges from $-90^\circ$ to 
$+90^\circ$, so that the output axis always points away from the input dipole.
The angles $\alpha$ and $\beta$ both start in the $x$-$y$-plane and describe the orientations $\vec{m}_1$ and 
$\vec{m}_2$ of the dipoles, which are represented by the thick red arrows.
Each dipole moment is perpendicular to its respective rotation axis.}
\label{fig:theory}
\end{figure}

The interaction between two point dipoles can be described following the textbook by Jackson.\cite{jackson}
The magnetic flux density of a dipole moment $\vec{m}_1$ sitting at the origin is given at a location $\vec{r}$
by
\begin{equation}
\label{eq:flux}
\vec{B}_1 (\vec{r}) = \frac{\mu_0}{4 \pi}\,\frac{3(\vec{m}_1\cdot\vec{r})\vec{r}-\vec{m}_1\,r^2}{r^5}
\end{equation}
If a second dipole $\vec{m}_2$ is located at $\vec{r}$, it has the potential energy 
$E = - \vec{m}_2 \cdot \vec{B}_1 (\vec{r})$. In order to model the setup described in section \ref{setup},
we use the coordinate system shown in Fig.~\ref{fig:theory}. The centers of the dipoles, with a distance 
$d$ between them, and their corresponding rotation axes lie in the horizontal $x$-$y$-plane and enclose 
with the $x$-axis the angles $\Theta$ and $\Phi$, respectively. In order to describe the experiments 
unambiguously, we choose the $z$-axis upwards in the vertical direction although gravity does not play a role.
The orientations of the two dipoles during their rotations are described by the angles $\alpha$ and $\beta$, 
respectively.

With these definitions and following the calculations by Sch\"onke,\cite{schonke2015b} 
the input and output dipole moments and the position vector can be written as
\begin{eqnarray}
\label{eq:vectors}
\vec{m}_1 & = & m_1\,\left( 
\begin{array}{c}
-\cos \alpha \, \sin \Theta \\
 \cos \alpha \, \cos \Theta \\
 \sin \alpha
\end{array}
\right) \, , \nonumber \\
\vec{m}_2 & = & m_2\,\left(
\begin{array}{c}
-\cos \beta \, \sin \Phi \\
 \cos \beta \, \cos \Phi \\
 \sin \beta
\end{array}
\right) \, , \quad
\vec{r} = d \,\left(
\begin{array}{c}
  1 \\
  0 \\
  0
\end{array}
\right) \, .
\end{eqnarray}
Inserting these expressions into Eq.~(\ref{eq:flux}), we find for the potential energy of the output
dipole $\vec{m}_2$ in the field of the input dipole $\vec{m}_1$
\begin{eqnarray}
\label{eq:energy}
E & & = \frac{\mu_0}{4 \pi}\,\frac{m_1\,m_2}{d^3} \cdot \\
  & &  \cdot \big(
        \sin \alpha \, \sin \beta + (\cos \Theta \, \cos \Phi - 2 \, \sin \Theta \, \sin \Phi) \, \cos \alpha \, \cos \beta
       \big) \, . \nonumber
\end{eqnarray}
The system is described by seven parameters: $m_1$, $m_2$ und $d$ define the strength of the
interaction, while the four angles $\Theta$, $\Phi$, $\alpha$ and $\beta$ define the type of 
rotation of the dipoles. For the further discussion, we rescale the energy to the dimensionless 
form $\tilde{E} = E / (\frac{\mu_0}{4 \pi}\,\frac{m_1\,m_2}{d^3})$.

\begin{figure}[b]
\centering
\includegraphics[width=1.05\columnwidth]{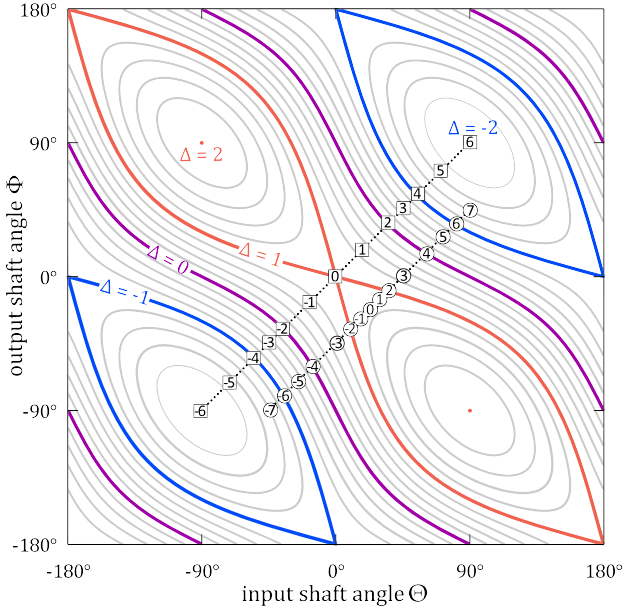}
\caption{Contour plot of the orientation index $\Delta$ in the $\Theta$--$\Phi$-plane (see Eq.~\ref{eq:delta}). 
The areas above $\Phi = 90^\circ$ and below $\Phi = - 90^\circ$ are steady continuations of the area 
within these borders, as $\Phi$ has been defined to range only from $-90^\circ$ to $+90^\circ$. 
The lines for $\Delta = 1$ (red), $\Delta = 0$ (violet) and $\Delta = -1$ (blue) are emphasized for better
readability.
The numbers from \protect\ensquare{-6} to \protect\ensquare{6} in the white squares and from 
\protect\encircle{-7} to \protect\encircle{7} in the white circles denote our 
measurement points, which will be discussed in section \ref{results}.}
\label{fig:delta}
\end{figure}

The orientation of the two rotation axes with respect to the line connecting the dipoles determines
the value of the \textit{(shaft) orientation index} $\Delta$,
\begin{equation}
\label{eq:delta}
\Delta = \cos \Theta \, \cos \Phi - 2 \, \sin \Theta \, \sin \Phi \, ,
\vspace*{-0.5ex}
\end{equation}
which is constrained to the interval between $-2$ and 2 and plotted in Fig.~\ref{fig:delta}.
The rescaled energy can then be written as
\begin{equation}
\label{eq:energy_dimless}
\tilde{E} = \sin \alpha \, \sin \beta + \Delta \, \cos \alpha \, \cos \beta \, .
\end{equation}
In our experiment, the angle $\alpha$ of the input dipole is prescribed by
a stepper motor, while the output dipole can rotate freely. 
The output is in an equilibrium position for
$\frac{\partial \tilde{E}}{\partial \beta} = \sin \alpha \, \cos \beta - \Delta \, \cos \alpha \, \sin \beta = 0$.
For $\alpha = \pm 90^\circ$ and $\beta = \pm 90^\circ$, this is fulfilled for any $\Delta$. These cases, 
where the dipoles are perpendicular to the $x$-$y$-plane, represent fix points for all configurations
independent of $\Delta$ and will be discussed later.
For $\alpha, \beta \ne \pm 90^\circ$, $\frac{\partial \tilde{E}}{\partial \beta} = 0$ is equivalent to
$\tan \alpha - \Delta \, \tan \beta = 0$.

For $\Delta \ne 0$, the equlibrium angle $\beta_\text{eq}$ of the output dipole can now be given
as a function of the input dipole angle $\alpha$ as
\begin{equation}
\label{eq:betaofalpha}
%\beta_\text{eq} = \arctan \left( \frac{1}{\Delta} \, \tan \alpha \right) + \, k \cdot 180^\circ \quad\text{with an integer $k$} .
\beta_\text{eq} = \arctan \left( \frac{1}{\Delta} \, \tan \alpha \right) + \, k \cdot 180^\circ \,\,\text{with integer $k$} .
\end{equation}

\begin{figure}[b]
\centering
\includegraphics[width=0.9\columnwidth]{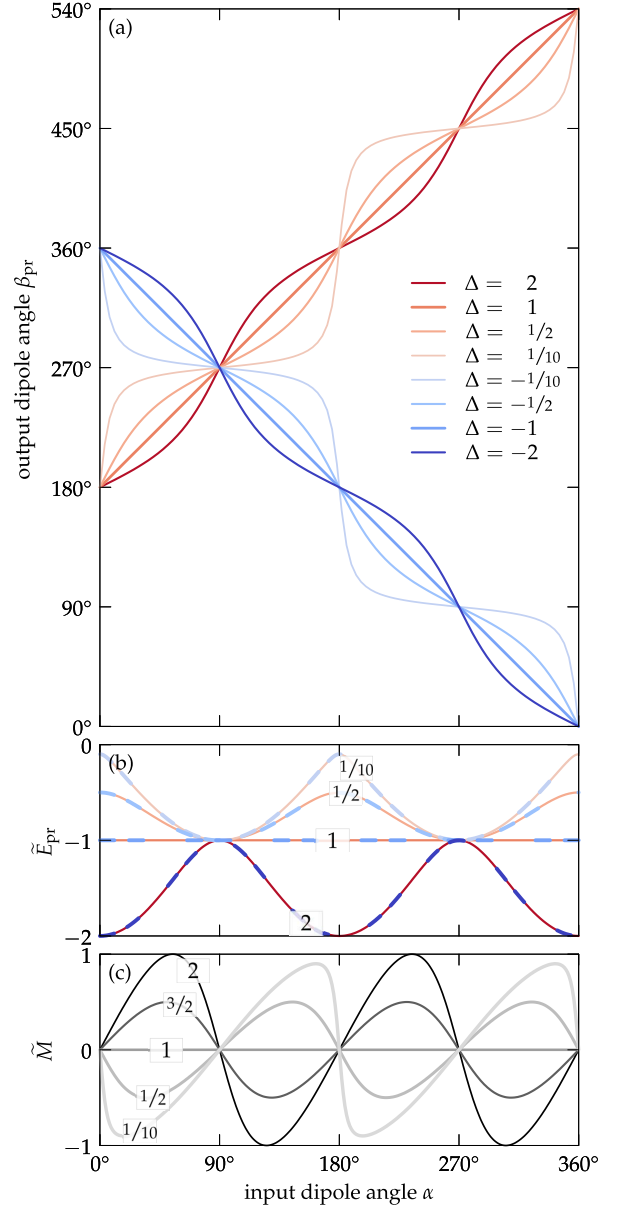}
\caption{The preferred angle $\beta_\text{pr}$ (a), the equilibrium enery $\tilde{E}_\text{pr}$ (b), and 
the cogging-torque $\tilde{M}$ (c) are shown as functions of the input dipole angle $\alpha$ for 
various values of $\Delta$. Cogging free couplings are realized 
for $\Delta = \pm 1$, leading to the two straight lines with slopes $\pm 1$ in (a), to a constant
$\tilde{E}_\text{eq} = -1$ in (b), and to $\tilde{M} = 0$ in (c). The numbers at the lines in (b) and (c)
are the respective values of $|\Delta|$.}
\label{fig:cogging}
\end{figure}

The system naturally chooses $k$ such that $\beta_\text{eq}$ minimizes $\tilde{E}$ locally 
($\frac{\partial^2 \tilde{E}}{\partial \beta^2} > 0$) yielding the prefered angle $\beta_\text{pr}$.
This condition restricts $k$ to 
$k = \frac{3}{2} + (\left[ \frac{\alpha}{180^\circ} \right] - \frac{1}{2}) \text{sgn} \Delta$,
where sgn is the signum function and the square brackets indicate convergent rounding to the nearest integer.
$\left[ \frac{\alpha}{180^\circ} \right]$ ensures the continuous extension of $\beta_\text{pr}$ 
outside the interval $[-180^\circ;+180^\circ]$ and may be omitted for small $\alpha$.
This energetically optimal angle is shown in Fig.~\ref{fig:cogging}(a) for some characteristical values 
of $\Delta$.

Obviously, the sign of $\Delta$ has the main influence on the behavior: For positive $\Delta$,
an increasing $\alpha$ results in an increasing $\beta_\text{pr}$, so that the two dipoles rotate in
the same direction, while a negative $\Delta$ leads to a counter-rotation. The absolute value
of $\Delta$ determines the smoothness of the interaction. For $\Delta = \pm 1$, 
$\beta_\text{pr}(\alpha)$ becomes a straight line with slope $\pm 1$, so that $\beta_\text{pr}$ follows $\alpha$
smoothly without a phase shift. The deviation from the straight line is determined by $\Delta$ and moreover,
it is the same for $\Delta$ and $1 / \Delta$, but on different sides of the line. 
This is demonstrated by $|\Delta| = 2$ and 
$|\Delta| = 0.5$ in Fig.~\ref{fig:cogging}(a). The closer $\Delta$ gets to 0, the more steplike
the function becomes.
The lines for the different configurations cross in one of the fix points mentioned above.

Inserting $\beta_\text{pr}$ into Eq.~(\ref{eq:energy_dimless}) yields the corresponding potential 
energy $\tilde{E}_\text{pr}$ as a function of $\alpha$ (see Fig.~\ref{fig:cogging}b for different $\Delta$).
Only the absolute value of $\Delta$ is important, so that the curves for $\Delta$ and
$-\Delta$ coincide. The meaning of the straight lines in Fig.~\ref{fig:cogging}(a) for 
$\Delta = \pm 1$ becomes clear now: the potential energy is constant and thus independent
of the angle of the driving dipole, so that no energy is needed to turn the second dipole: 
a cogging-free clutch!\cite{schonke2015b} For $\Delta \neq \pm 1$ this is not the case, and the amplitude of 
$\tilde{E}_\text{eq}$ in Fig.~\ref{fig:cogging}(b) represents the energy needed during half a turn.

Differentiating the potential energy in Eq.~(\ref{eq:energy}) with respect to $\alpha$, we find
the cogging-torque $M$, an important variable for the motor. The dimensionless cogging torque 
$\tilde{M} = \frac{\partial \tilde{E}}{\partial \alpha}$ is shown in Fig.~\ref{fig:cogging}(c). 
Again, the special meaning of the fix points is emphasized: no cogging torque is exerted on the
second dipole.

\begin{figure}[b]
\centering
\includegraphics[width=\columnwidth]{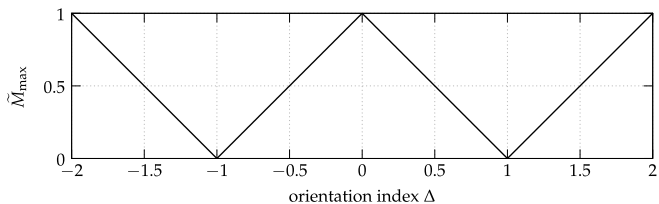}
\caption{The torque $\tilde{M}_\text{max}$ that is needed to rotate the driving dipole by half a turn
is shown as a function of the orientation index $\Delta$. 
For $\Delta = \pm 1$, this torque is zero while for any other values of $\Delta$, a finite torque is required.}
\label{fig:torque}
\end{figure}

In order to rotate the driving dipole by half a turn, a finite torque must be exerted
which is given by the maximum $\tilde{M}_\text{max}$ of the respective curve. This shows a 
surprisingly simple dependence on $\Delta$:
\begin{equation}
\label{eq:cogg_delta}
\tilde{M}_\text{max} = \big| 1 - | \Delta | \big|
\end{equation}
and is plotted in Fig.~\ref{fig:torque}. Again, the special cases $\Delta = \pm 1$ 
are obvious, leading to vanishing torques. In the $\Delta$-landscape plot of 
Fig.~\ref{fig:delta}, the contour lines for $\Delta = \pm 1$ are emphasized, depicting the pairs of
$\Theta$ and $\Phi$ for vanishing torques. Since $\Delta(\Theta, \Phi)$
is a continuous function, there is always a line with $\Delta = 0$ between those two lines. Thus, by
crossing the line $\Delta = 0$ through variation of $\Theta$ or $\Phi$, the sense of rotation
of the output dipole with respect to the driving dipole changes.

\begin{figure}[t]
\centering
\includegraphics[width=\columnwidth]{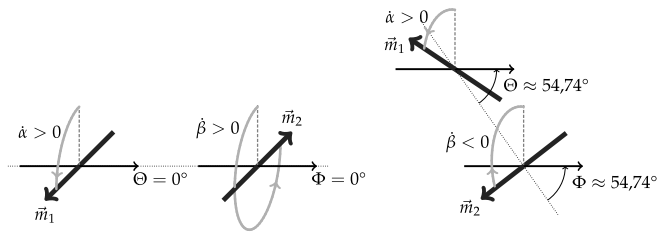}
\caption{The two cogging free couplings for $\Theta = \Phi$. While the left part 
represents the trivial coupling with co-rotating dipoles like in a magnetic stirrer 
($\Theta = \Phi = 0$, $\Delta = 1$), the right image represents the nontrivial coupling 
with counter-rotating dipoles ($\Theta = \Phi \approx 54.74^\circ$, $\Delta = -1$). The
grey arrows indicate the direction of rotation of the respective dipoles.}
\label{fig:rotation}
\end{figure}

Let us now assume that the input and output shafts are parallel, so that $\Theta = \Phi$. This situation
corresponds to the black diagonal line through the center of Fig.~\ref{fig:delta}. The point
in the middle, $\Theta = \Phi = 0$, belongs to a co-axial setup with $\Delta = 1$, so that the 
two dipoles rotate cogging free in the same direction (see Fig.~\ref{fig:rotation}, left). 
By parallel shifting one of the two shafts, we move along the black line in Fig.~\ref{fig:delta}, 
e.g.\ upwards to the right through $\Delta = 0$ (maximal cogging torque), thus changing the sense of rotation,
to the point with $\Delta = -1$. This point is reached for $\Theta = \Phi \approx 54.74^\circ$ and
again corresponds to a cogging free rotation, but now with the two dipoles rotating in 
opposite directions (see Fig.~\ref{fig:rotation}, right). Thus we can shift gears by shifting the two axes.

Couplings with a finite but fixed angle between the two shafts correspond to lines in 
Fig.~\ref{fig:delta} which are parallel to the one just discussed. An example for
$\Theta = \Phi - 45^\circ$ is shown by the lower black line, and again
we can switch from co- to counter-rotating dipoles by shifting the axes.

\section{Experimental Results and Discussion}
\label{results}	
Two measurement series have been performed: In the first series, the input and output shafts are parallel 
($\Phi = \Theta$), while in the second series, the two rotation axes enclose an angle of $45^\circ$ 
($\Phi = \Theta - 45^\circ$). The input shaft is turned with one round per minute, the slowest motion
possible for the stepper motor software. After each motion we wait for the output shaft to come to a hold 
before measuring the output angle $\beta$. In this way, we try to realize the limit of vanishing speed of 
the drives. During the experiments, the internal friction in the bearings is the only load on the output 
shaft. In both series, we consider situations with minimal, with maximal and with intermediate cogging torques.

\subsection{Parallel rotation axes: $\Phi = \Theta$}
\label{parallel}	
In the simplest case of the dipole-dipole coupling, the two axes are parallel.
The important parameters for this set of experiments are summarized in Tab.~\ref{tab:parallel}.
$\Theta$ and $\Phi$ can be adjusted to approximately $1^\circ$, leading to an average uncertainty 
for $\Delta$ of about 0.04 which can rise to 0.08 near $\Delta = 0$. Listed in Tab.~\ref{tab:parallel}
are the values $\Theta_\text{tar}$ and $\Phi_\text{tar}$, that we target in the experiment, together
with the corresponding $\Delta_\text{tar}$.

\begin{table}[b]
\centering
\caption{Parameters for parallal input and output axes ($\Phi = \Theta$) for the various measurement 
configurations \protect\ensquare{-6} to \protect\ensquare{6}. The targeted output and input shaft angles $\Phi_\text{tar}$ and 
$\Theta_\text{tar}$ are adjusted manually, resulting in $\Delta_\text{tar}$. $\Delta_\text{fit}$ 
and $\alpha_0$ are results of a fit of Eq.~(\ref{eq:betaofalpha}) to the data points.}
\begin{ruledtabular}
\begin{tabular}{c c c c c c}
configuration & $\Phi_\text{tar}$ ($^\circ$) & $\Theta_\text{tar}$ ($^\circ$) & $\Delta_\text{tar}$ & $\Delta_\text{fit}$ & $\alpha_0$  ($^\circ$)\\ \hline	
\ensquare{-6} & $-90.00$          & $-90.00$            & $-2.000$            & $-2.025$            & $6.17$     \\
\ensquare{-5} & $-72.37$          & $-72.37$            & $-1.725$            & $-1.678$            & $3.60$     \\
\ensquare{-4} & $-54.74$          & $-54.74$            & $-1.000$            & $-0.926$            & $-3.59$    \\
\ensquare{-3} & $-45.00$          & $-45.00$            & $-0.500$            & $-0.457$            & $11.35$    \\
\ensquare{-2} & $-35.26$          & $-35.26$            & $0.001$             & $0.085$             & $9.74$     \\
\ensquare{-1} & $-22.50$          & $-22.50$            & $0.561$             & $0.610$             & $8.54$     \\
\ensquare{0}  & $0.00$            & $0.00$              & $1.000$             & $0.998$             & $8.10$     \\
\ensquare{1}  & $22.50$           & $22.50$             & $0.561$             & $0.516$             & $8.44$     \\
\ensquare{2}  & $35.26$           & $35.26$             & $-0.001$            & $-0.058$            & $8.99$     \\
\ensquare{3}  & $45.00$           & $45.00$             & $-0.500$            & $-0.588$            & $8.99$     \\
\ensquare{4}  & $54.74$           & $54.74$             & $-1.000$            & $-1.133$            & $8.91$     \\
\ensquare{5}  & $72.37$           & $72.37$             & $-1.725$            & $-1.761$            & $9.00$     \\
\ensquare{6}  & $90.00$           & $90.00$             & $-2.000$            & $-2.028$            & $-3.31$    \\
\end{tabular}
\end{ruledtabular}
\label{tab:parallel}
\end{table}

The experimental $\Delta$-values are shown in Fig.~\ref{fig:delta} by the upper dashed line,
which runs diagonally through the center of the graph, and where the white squares with the numbers
\ensquare{-6} to \ensquare{6} represent our measurement configurations. Numbers with the same 
absolute value lead to the same $\Delta$, so that similar properties are expected for the respective 
couplings. Our measurements are confined to values of $-90^\circ \le \Theta \le 90^\circ$, as a 
configuration with  $|\Theta| > 90^\circ$ always corresponds to one with $|\Theta| < 90^\circ$ 
except for the sense of rotation.

\begin{figure}[b]
\centering
\includegraphics[width=\columnwidth]{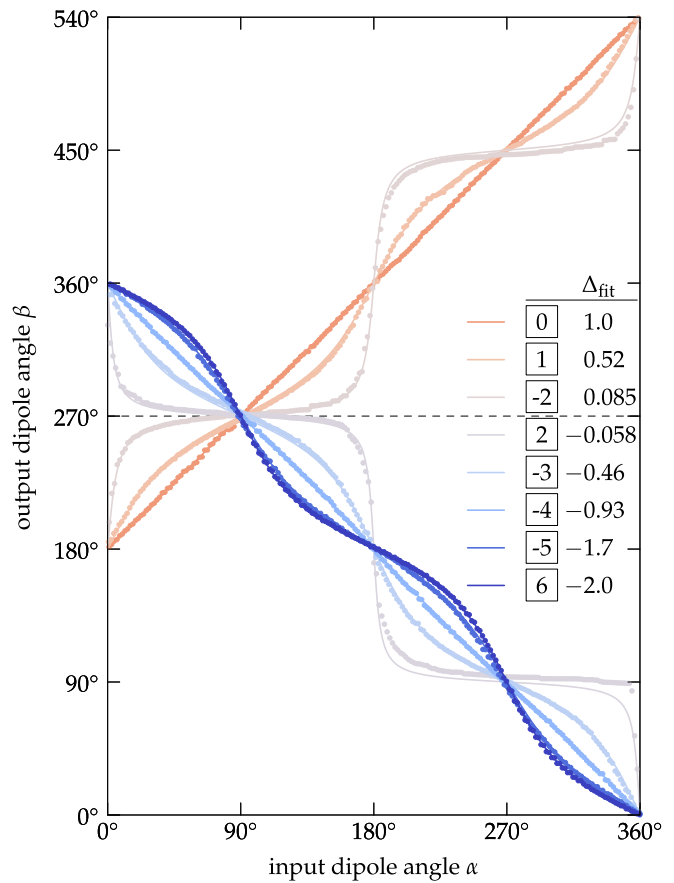}
\caption{The measured output angle $\beta$ as a function of the input angle $\alpha$ for the measurement
configurations \protect\ensquare{0}, \protect\ensquare{1}, \protect\ensquare{-2}, \protect\ensquare{2},
\protect\ensquare{-3}, \protect\ensquare{-4}, \protect\ensquare{-5} and \protect\ensquare{6} with the
corresponding $\Delta_\text{fit}$ in the inset from top to bottom.}
\label{fig:results1}
\end{figure}

The measurements directly yield the dependence of the output dipole angle $\beta$ on the input
dipole angle $\alpha$, i.e.\ the curves $\beta(\alpha)$ which are shown in Fig.~\ref{fig:results1}.
Co-rotating dipoles ($\Delta > 0$) are represented by the branch that extends to the right upwards,
while counter-rotating dipoles ($\Delta < 0$) are represented by the branch that extends to the right 
downwards. We find cogging free couplings for three of the configurations:
the measurements for configuration \ensquare{0} with $\Delta = 1$ and for the two configurations 
\ensquare{-4} (and \ensquare{4}, not shown in Fig.~\ref{fig:results1}) with $\Delta = -1$ are in 
good approximation linear in the $\beta(\alpha)$-graph. The coupling \ensquare{0} shows a trivial 
rotational symmetry, where the two rotation axes are not only parallel but also co-axial, as shown 
in Fig.~\ref{fig:rotation}, left. The couplings for configurations \ensquare{-4} (and \ensquare{4}), 
on the other hand, are not quite as obvious. The axes are still parallel, but laterally shifted as 
shown in Fig.~\ref{fig:rotation}, right, leading to a non-trivial cogging free counter-rotation. 
These results are in agreement with the theory.

\begin{figure}[b]
\centering
\includegraphics[width=\columnwidth]{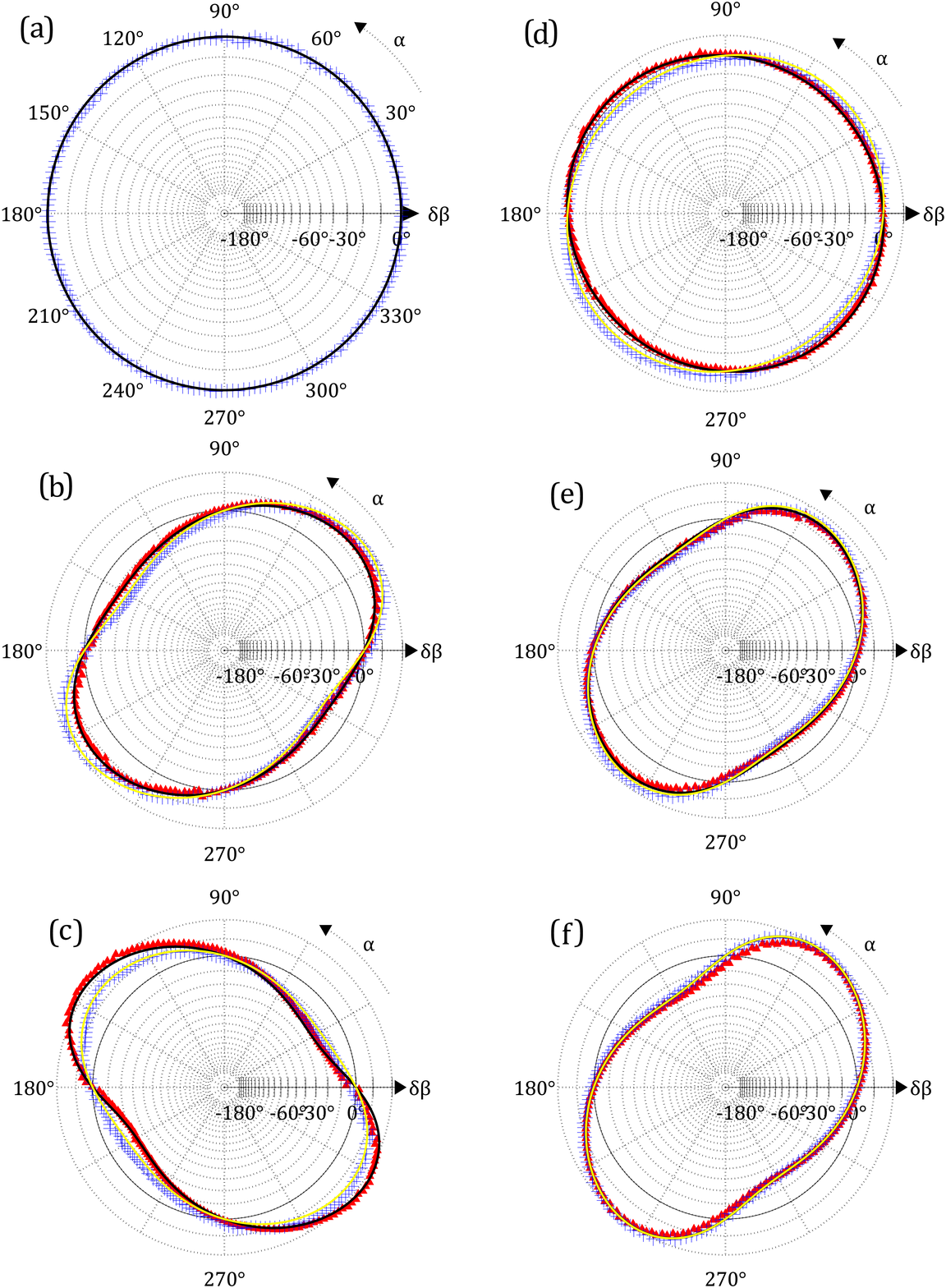}
\caption{Phase jitter $\delta \beta$ as a function of the input dipole angle $\alpha$ for the measurements 
with parallel rotation axes. 
(a) Measurement configuration \protect\ensquare{0} with $\Theta \approx \Phi \approx 0^\circ$.
(b) Configurations \protect\ensquare{-1} with $\Theta \approx \Phi \approx -22.5^\circ$ and \protect\ensquare{1}
with $\Theta \approx \Phi \approx 22.5^\circ$.
(c) Configurations \protect\ensquare{-3} with $\Theta \approx \Phi \approx -45^\circ$ and \protect\ensquare{3}
with $\Theta \approx \Phi \approx 45^\circ$.
(d) Configurations \protect\ensquare{-4} with $\Theta \approx \Phi \approx -55^\circ$ and \protect\ensquare{4}
with $\Theta \approx \Phi \approx 55^\circ$.
(e) Configurations \protect\ensquare{-5} with $\Theta \approx \Phi \approx -72^\circ$ and \protect\ensquare{5}
with $\Theta \approx \Phi \approx 72^\circ$.
(f) Configurations \protect\ensquare{-6} with $\Theta \approx \Phi \approx -90^\circ$ and \protect\ensquare{6}
with $\Theta \approx \Phi \approx 90^\circ$.
The red triangles are for configuration with negative numbers, while the blue squares are for 
positive numbers. The black lines are the respective fits for negative numbers, while the yellow lines
are the fits for positive numbers.}
\label{fig:results2}
\end{figure}

The further $\Delta$ deviates from $\pm 1$, the more the $\beta(\alpha)$-curve deviates from
the linear dependence. For $\Delta$ close to 0, the dependence becomes almost steplike 
(measurement configurations \ensquare{-2} and \ensquare{2}). When $\alpha$ increases from 
$90^\circ$ to $180^\circ$, for most of the turn $\beta$ hardly changes and then suddenly jumps 
to the new equilibrium values. This corresponds to the fact that for $\Delta = 0$, the coupling 
changes from co- to counter-rotation. For half of the cycle, $\beta$ lags extremely behind $\alpha$, 
while for the other half, $\beta$ is far ahead of $\alpha$.

Together with the experimental data points, fits of Eq.~(\ref{eq:betaofalpha}) to the data are also 
shown in Fig.~\ref{fig:results1}. As the starting point of $\alpha$ is not perfectly known due to 
static friction of the shafts in the bearings and also due to uncertainties while adjusting the setup 
to a new $\Delta$, an offset $\alpha_0$ is assumed in the fits. Therefore, our free fit parameters are 
$\Delta$ and $\alpha_0$. Our measurement data are in very good agreement with the theoretical 
predictions with the biggest deviations for configurations with the maximal cogging torque, i.e.\ 
near $\Delta = 0$. In Tab.~\ref{tab:parallel}, the results of the fit procedure for $\Delta_\text{fit}$ 
and $\alpha_0$ are also given. $\Delta_\text{fit}$ is always close to the value $\Delta_\text{tar}$ 
assumed in the experiment with the biggest relative deviations near $\Delta = 0$.

The phase jitter is plotted in polar coordinates in Fig.~\ref{fig:results2} for the measurement configurations
\ensquare{0}, \ensquare{-1}\&\ensquare{1}, \ensquare{-3}\&\ensquare{3}, \ensquare{-4}\&\ensquare{4},
\ensquare{-5}\&\ensquare{5}, and \ensquare{-6}\&\ensquare{6}. Again, the angular coordinate shows the 
input dipole angle $\alpha$ and the radial coordinate the phase jitter $\delta \beta$, i.e.\ how far the output 
is ahead (positive values of $\delta \beta$) or lags behind (negative values of $\delta \beta$) the input.
The data for similar values of $\Delta$, which should lead to similar results, are always plotted in the 
same diagram together with the respective fits. The qualitative agreement ist obvious, albeit some bigger 
deviations can be seen for configurations \ensquare{-1}\&\ensquare{1}, \ensquare{-3}\&\ensquare{3}, and 
\ensquare{-4}\&\ensquare{4} (see Fig.~\ref{fig:results2}~b, c and d). 
Here, the values for the fitted $\Delta$ of the complementary data show larger differences.

\begin{figure}[b]
\centering
\includegraphics[width=\columnwidth]{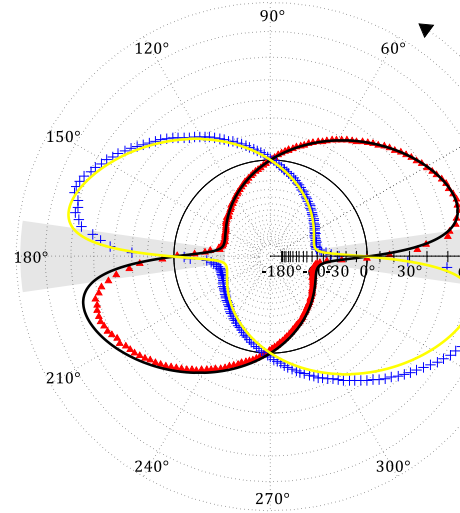}
\caption{Phase jitter $\delta \beta$ as a function of the input dipole angle $\alpha$ for measurements near
$\Delta = 0$, where maximal cogging ist expected. The red triangles are for measurement configuration
\protect\ensquare{-2} ($\Theta \approx \Phi \approx -35^\circ$) while the blue squares are for configuration
\protect\ensquare{2} ($\Theta \approx \Phi \approx 35^\circ$). The yellow and the black lines are the respective
fits. Within the shaded areas, the output jumps from a maximal value of $\delta\beta$ to its minimal value.}
\label{fig:results1b}
\end{figure}

The biggest deviation from a cogging free coupling is expected near $\Delta = 0$. These results (measurement
configurations \ensquare{-2}\&\ensquare{2}) are shown in Fig.~\ref{fig:results1b}, where the red points 
represent a slightly positive $\Delta$ and the blue points a slightly negative $\Delta$. It can be seen, 
that a slight change of $\Delta$ leads to a dramatic change in the qualitative behaviour, as the dipoles 
change from co-rotating to counter-rotating when crossing $\Delta = 0$. Looking at the blue data points, during
one half turn of the input dipole, $\delta \beta$ steadily increases until it reaches a maximal value near
$\delta \beta \approx 90^\circ$. This is due to the fact that during this phase the output dipole hardly
moves although the input dipole keeps rotating. The output lags more and more behind until the input
reaches a threshold near $\alpha \approx 0^\circ$ or near $180^\circ$, when the torque transmission of
the coupling is maximal and the output jumps from the maximal value of $\delta \beta$ to its minimal value
(indicated by the shaded areas in Fig.~\ref{fig:results1b}). This jumping behavior can be attributed to the
fact that the stepper motor does not turn continuously but rather has a finite step width, so that the output
always has to overcome the static friction.

The theoretical model fits the data very well, leading to a value for $\Delta_\text{fit}$ slightly different 
from the experimentally assumed $\Delta_\text{exp}$. This can be explained by the uncertainty of the 
positioning of the shafts. In some cases, however, the fit does not work as well, as can be seen especially for
the values of $\Delta$ near zero in Fig.~\ref{fig:results1b}. One reason might be the friction in the bearings
and the finiteness of the motor steps. Moreover, one should not forget, that the magnets
are macroscopic industrial objects and that they may not be exact point dipoles (see appendix \ref{app:dipole}). 
Finally, it is also possible, that the dipoles are not perfectly perpendicular to the respective axis, which may explain 
a slightly different behaviour for complementary situations.

\begin{table}[b]
\centering
\caption{Parameters for non-parallal input and output axes ($\Phi = \Theta - 45^\circ$) for the various 
measurement configurations \protect\encircle{-7} to \protect\encircle{7}. Further explanations see Tab.~\ref{tab:parallel}.}
\begin{ruledtabular}
\begin{tabular}{c c c c c c}
configuration & $\Phi_\text{tar}$ ($^\circ$) & $\Theta_\text{tar}$ ($^\circ$) & $\Delta_\text{tar}$ & $\Delta_\text{fit}$ & $\alpha_0$ ($^\circ$)\\ \hline	
\encircle{-7} & $-90.00$          & $-45.00$            & $-1.414$            & $-1.399$            & $7.20$     \\
\encircle{-6} & $-80.27$          & $-35.27$            & $-1.000$            & $-1.008$            & $6.79$     \\
\encircle{-5} & $-70.47$          & $-25.47$            & $-0.509$            & $-0.500$            & $7.19$     \\
\encircle{-4} & $-60.68$          & $-15.66$            & $0.001$             & $0.129$             & $10.10$    \\
\encircle{-3} & $-45.00$          & $0.00$              & $0.707$             & $0.735$             & $4.32$     \\
\encircle{-2} & $-35.26$          & $9.74$              & $1.000$             & $1.007$             & $3.93$     \\
\encircle{-1} & $-28.82$          & $16.12$             & $1.109$             & $1.119$             & $3.75$     \\
\encircle{0}  & $-22.50$          & $22.50$             & $1.146$             & $1.143$             & $3.97$     \\
\encircle{1}  & $-16.12$          & $28.82$             & $1.109$             & $1.093$             & $3.94$     \\
\encircle{2}  & $-9.74$           & $35.26$             & $1.000$             & $0.977$             & $3.59$     \\
\encircle{3}  & $0.00$            & $45.00$             & $0.707$             & $0.688$             & $3.89$     \\
\encircle{4}  & $15.68$           & $60.65$             & $0.001$             & $0.132$             & $5.39$     \\
\encircle{5}  & $25.47$           & $70.47$             & $-0.509$            & $-0.600$            & $5.94$     \\
\encircle{6}  & $35.27$           & $80.27$             & $-1.000$            & $-1.039$            & $5.39$     \\
\encircle{7}  & $45.00$           & $90.00$             & $-1.414$            & $-1.469$            & $5.30$     \\
\end{tabular}
\end{ruledtabular}
\label{tab:nonparallel}
\end{table}

\subsection{Non-parallel rotation axes: $\Phi \neq \Theta$}
\label{nonparallel}	
In a second series of experiments, we consider two rotation axes which are not parallel, but rather enclose
an angle of $45^\circ$: $\Phi = \Theta - 45^\circ$. The important parameters for this set of experiments are 
summarized in Tab.~\ref{tab:nonparallel}. The corresponding $\Delta$-values are shown in Fig.~\ref{fig:delta} 
by the lower dashed line, where our measurement configurations are represented by the white circles with the 
numbers \encircle{-7} to \encircle{7}. Again, numbers with the  same absolute value have the same $\Delta$. 

\begin{figure}[b]
\centering
\includegraphics[width=\columnwidth]{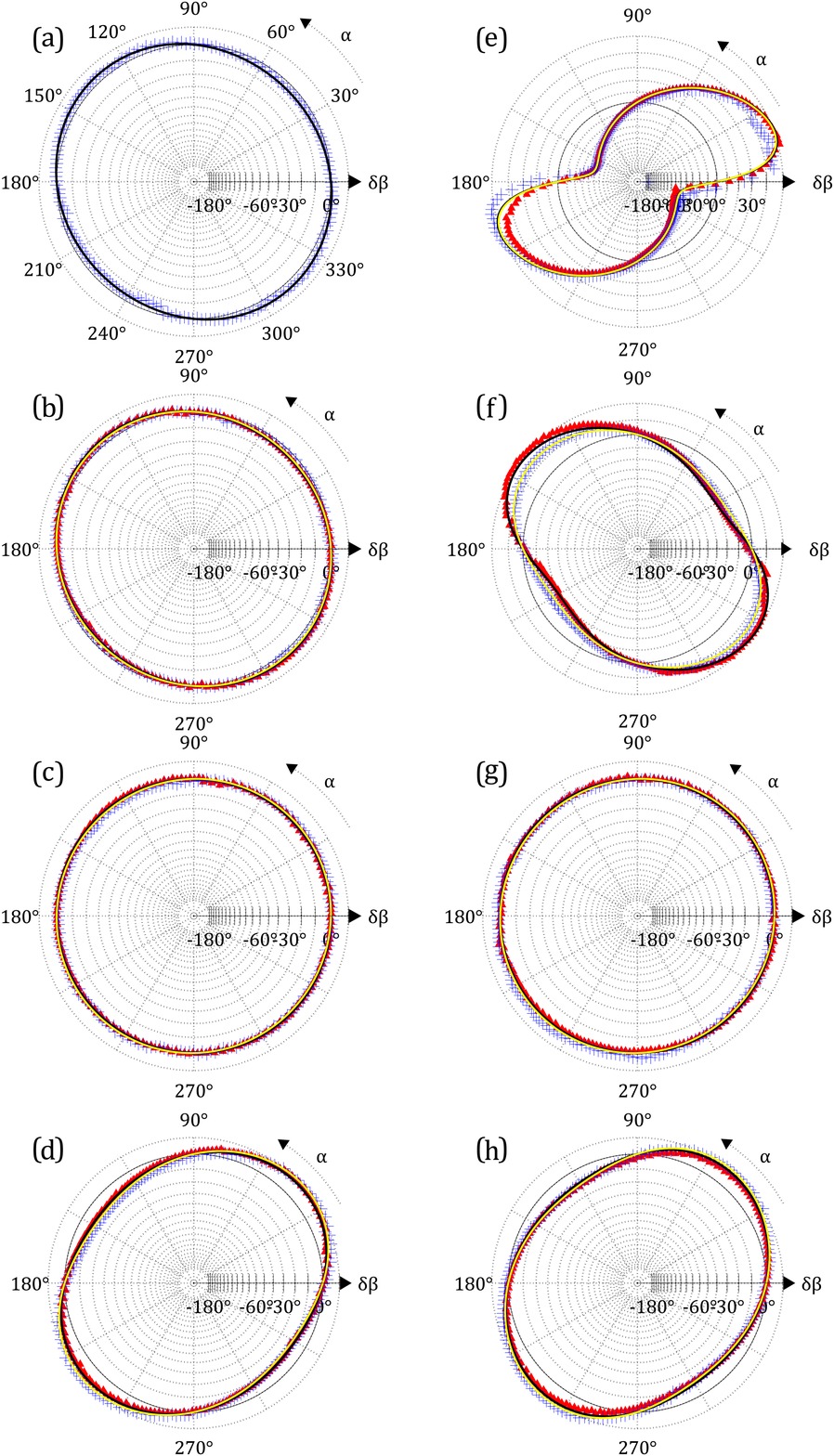}
\caption{Phase jitter $\delta \beta$ as a function of the input dipole angle $\alpha$ for the measurements 
with non-parallel rotation axes. 
(a) Measurement configuration \protect\encircle{0} with $\Theta \approx 22.5^\circ$, $\Phi \approx -22.5^\circ$.
(b)~Configurations \protect\encircle{-1} with $\Theta \approx 16^\circ$, $\Phi \approx -29^\circ$ and \protect\encircle{1}
with $\Theta \approx 29^\circ$, $\Phi \approx -16^\circ$.
(c) Configurations \protect\encircle{-2} with $\Theta \approx 10^\circ$, $\Phi \approx -35^\circ$ and \protect\encircle{2}
with $\Theta \approx 35^\circ$, $\Phi \approx -10^\circ$.
(d) Configurations \protect\encircle{-3} with $\Theta \approx 0^\circ$, $\Phi \approx -45^\circ$ and \protect\encircle{3}
with $\Theta \approx 45^\circ$, $\Phi \approx 0^\circ$.
(e) Configurations \protect\encircle{-4} with $\Theta \approx -16^\circ$, $\Phi \approx -61^\circ$ and \protect\encircle{4}
with $\Theta \approx 61^\circ$, $\Phi \approx 16^\circ$.
(f) Configurations \protect\encircle{-5} with $\Theta \approx -25^\circ$, $\Phi \approx -70^\circ$ and \protect\encircle{5}
with $\Theta \approx 70^\circ$, $\Phi \approx 25^\circ$.
(g) Configurations \protect\encircle{-6} with $\Theta \approx -35^\circ$, $\Phi \approx -80^\circ$ and \protect\encircle{6}
with $\Theta \approx 80^\circ$, $\Phi \approx 35^\circ$.
(h) Configurations \protect\encircle{-7} with $\Theta \approx -45^\circ$, $\Phi \approx -90^\circ$ and \protect\encircle{7}
with $\Theta \approx 90^\circ$, $\Phi \approx 45^\circ$.
The red triangles are for configuration with negative numbers, while the blue squares are for 
positive numbers. The black lines are the respective fits for negative numbers, while the yellow lines
are the fits for positive numbers.}
\label{fig:results2b}
\end{figure}

The phase jitter is plotted in polar diagrams in Fig.~\ref{fig:results2b} together with their fits. Again,
there is a good qualitative agreement between theory and experiment as well as for configurations with
the same $\Delta$. The almost perfectly circular graphs for measurement configurations \encircle{-2}\&\encircle{2} 
and \encircle{-6}\&\encircle{6} show a quantitative agreement with the theory 
(see Fig.~\ref{fig:results2b}c and g). Here, $|\Delta| = 1$, so that cogging free couplings are expected. 
Both situations represent non-trivial couplings. For $\Phi \approx -10^\circ$ 
and $\Theta \approx 35^\circ$ (configuration \encircle{2}) as well as $\Phi \approx -35^\circ$ and 
$\Theta \approx 10^\circ$ (configuration \encircle{-2}) we have $\Delta = 1$, so that the dipoles
are co-rotating. For $\Phi \approx 35^\circ$ and $\Theta \approx 80^\circ$ (configuration \encircle{6}) 
and $\Phi \approx -80^\circ$ and $\Theta \approx -35^\circ$ (configuration \encircle{-6}), on the other 
hand, $\Delta = -1$, leading to counter-rotating dipoles. The other extreme of maximal cogging torque 
($\Delta \approx 0$) is found for our measurement configurations \encircle{4} ($\Phi \approx 16^\circ$ 
and $\Theta \approx 61^\circ$) and \encircle{-4} ($\Phi \approx -61^\circ$ and $\Theta \approx -16^\circ$) 
and are shown in Fig.~\ref{fig:results2b}(e).

\begin{figure}[b]
\centering
\includegraphics[width=\columnwidth]{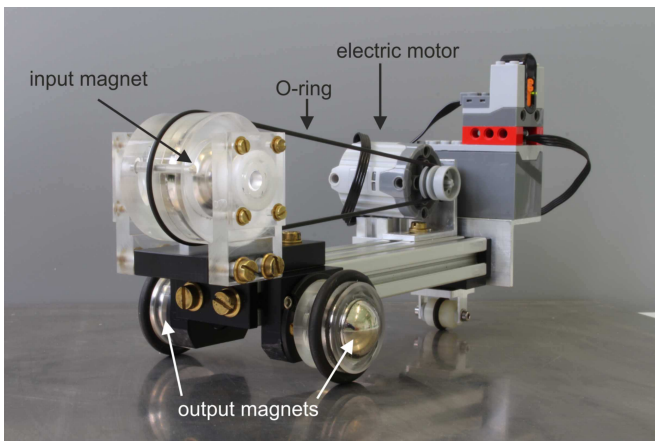}
\caption{Toy car using a cogging free coupling with three instead of two magnets.\cite{schonke2015b,schonke2016}
The big input magnet is coupled to an electric motor by an O-ring, so that it can rotate with a 
constant speed. This magnet drives the two identical, but smaller output magnets which are in such 
positions, that the coupling is cogging free. The dipole moments of all three magnets are 
perpendicular to their respective rotation axes. A video of the moving car is avalable at 
\protect\url{http://www.staff.uni-bayreuth.de/~bt190039/download/AJP/car_movie.mp4}.}
\label{fig:car}
\end{figure}

\section{Conclusion and Outlook}
\label{conclusion}
In summary, these experiments demonstrate the crossover from trivial to non-trivial cogging free magnetic 
gears. The experiments can be done with commercially available spherical magnets. Their external magnetic 
field is in very good approximation that of an idealized point dipole. Therefore, an elementary 
theoretical model based on static interactions of point dipoles seems to be the adequate description of 
the experiments. To broaden the outlook towards technical application, we have built the toy model shown in 
Fig.~\ref{fig:car}, which makes use of the nontrivial coupling described in this paper. It shows the 
striking feature that the motor rotates in the opposite direction than the driving wheels which makes 
it particularly useful for classroom demonstrations. Technical applications of this kind of coupling 
are foreseeable for miniaturized machines where mechanical gears might be unwieldy.

\appendix*   % Omit the * if there's more than one appendix.
\label{app:dipole}

\section{Dipole approximation for the spherical magnets}
\begin{figure}[b]
\centering
\includegraphics[width=\columnwidth]{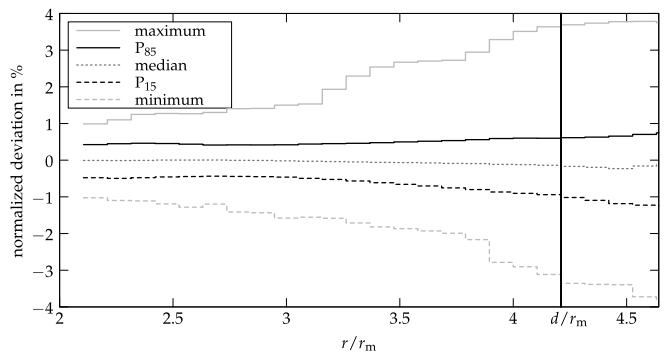}
\caption{
Characteristic percentiles for normalized deviation histograms as functions of the distance $r$ 
from the center of the sphere. $r_\text{m}$ is the radius of the sphere and the vertical line 
marks the distance $d$ that was realized between the two spheres in the magnetic clutch experiments 
presented in this paper. No residuals were found below the ``minimum'' and above the ``maximum'' 
line, while 50\% are above and below the ``median'' line. The center 70\% of the residuals, which 
indicate the standard deviation when approximating the histograms as gaussian distributions, lie
between the P$_{15}$ and the P$_{85}$ lines. Therefore, the typical deviation from the ideal dipole 
approximation in our experiment is typically within 1\%.
}
\label{fig:quality}
\end{figure}

In order to find out how well the field of the magnets we used can be modelled by an ideal point dipole,
we performed measurements of the field of the input magnet.\cite{volkel2016}
For these experiments, the output shaft is 
replaced by a hall probe (MMT-6J02-VH with 450 Gaussmeter, Lake Shore) mounted on an $X$-$Y$ table which 
is oriented such that the probe scans in a plane parallel to the table on which the experiments are performed.
The scanning plane is chosen as closely to the center of the input magnet as the mounting allows and lies 
$20\,$mm above the magnet. We sample the magnetic field at $100 \times 100$ points within a rectangle of 
$62.5\,\text{mm} \times 75\,\text{mm}$. At each point, the component of the magnetic field parallel 
to the measurement plane and perpendicular to the input shaft is measured. After each scan the input 
shaft is rotated by $4.5^{\circ}$, thus allowing to average over all possible orientations between 
the dipole and the measurement plane.

An ideal dipole field is fitted to each scan and the residuals are calculated for each point. These 
residuals are normalized by the maximal absolute value of the magnetic field at the respective distance.
The resulting normalized deviation from a point dipole approximation is then evaluated statistically
yielding a histogram for each distance. Fig.~\ref{fig:quality} shows five characteristic percentiles 
of each histogram  as functions of the distance from the center of the sphere. From this we can deduct 
that the normalized deviation from an ideal point dipole is approximately constant for the distances  
investigated here and in our experiments it is typically within 1\%.

\begin{acknowledgments}

We gratefully acknowledge Johannes Sch\"onke for enlightening discussions and Klaus Oetter for his
skillful help by designing the experimental setup. The experiments were supported by Deutsche 
Forschungsgemeinschaft (DFG) through grant RE 588/20-1.

\end{acknowledgments}


\begin{thebibliography}{99}
% The numeral (here 99) in curly braces is nominally the number of entries in
% the bibliography. It's supposed to affect the amount of space around the
% numerical labels, so only the number of digits should matter--and even that
% seems to make no discernible difference.

\bibitem{hirschfelder1964} 
Joseph O.\ Hirschfelder, Charles F.\ Curtiss, and R.\ Byron Bird, 
\textit{Molecular Theory of Gases and Liquids}, 8th edition (John Wiley \& Sons Inc., 1964). 

\bibitem{adams2007} 
Al Adams, 
``Spherical Rare-Earth Magnets in Introductory Physics,'' 
Phys. Teach. \textbf{45}, 409--415 (2007).
%http://aapt.scitation.org/doi/pdf/10.1119/1.2783147 

\bibitem{griffiths1992} 
David J.\ Griffiths, 
``Dipoles at rest,'' 
Am. J. Phys. \textbf{60}, 979--987 (1992).
%http://aapt.scitation.org/doi/pdf/10.1119/1.17001 

\bibitem{edwards2017} 
Boyd F.\ Edwards, D.\ M.\ Riffe, Jeong-Young Ji, and William A.\ Booth, 
``Interactions between uniformly magnetized spheres,'' 
Am. J. Phys. \textbf{85}, 130--134 (2017).
%http://aapt.scitation.org/doi/pdf/10.1119/1.4973409). 
 
\bibitem{luttinger1946} 
J.\ M.\ Luttinger and L.\ Tisza, 
``Theory of Dipole Interaction in Crystals,'' 
Phys. Rev. \textbf{70}, 954--964 (1946).

\bibitem{belobrov1983} 
Peter J.\ Belobrov, R.\ S.\ Gekht, and V.\ A.\ Ignatchenko,
``Ground state in systems with dipole interaction,''
Sov. Phys. JETP \textbf{57}, 636--642 (1983). 

\bibitem{vandewalle2014} 
N.\ Vandewalle and S.\ Dorbolo, 
``Magnetic ghosts and monopoles,''  
New J. Phys. \textbf{16}, 013050 (2014). 

\bibitem{schonke2015a} 
Johannes Sch\"onke, Tobias M. Schneider, and Ingo Rehberg,
``Infinite geometric frustration in a cubic dipole cluster,'' 
Phys. Rev. B \textbf{91}, 020410-1--020410-5 (2015).  

\bibitem{messina2014} 
Ren\'e Messina, Lara Abou Khalil, and Igor Stankovic, 
``Self-assembly of magnetic balls: From chains to tubes,'' 
Phys. Rev. E \textbf{89}, 011202(R) (2014).

\bibitem{vella2014} 
Dominic Vella, Emmanuel du Pontavice, Cameron L.\ Hall, and Alain Goriely, 
``The magneto-elastica: from self-buckling to self-assembly,'' 
Proc. R. Soc. A \textbf{470}, 20130609 (2014).
% http://dx.doi.org/10.1098/rspa.2013.0609 
  
\bibitem{schonke2017} 
Johannes Sch\"onke and Eliot Fried, 
``Stability of vertical magnetic chains,'' 
Proc. R. Soc. A \textbf{473} 20160703 (2017).
%http://dx.doi.org/10.1098/rspa.2016.0703 

\bibitem{sprott2006} 
Julien Clinton Sprott,
\textit{Physics Demonstrations: A Sourcebook for Teachers of Physics}, 
(The University of Wisconsin Press, Madison Wisconsin, 2006).

\bibitem{pollack1997} 
Gerald L.\ Pollack and Daniel R.\ Stump,
``Two magnets oscillating in each other's fields,'' 
Can. J. Phys. \textbf{75}, 313--324 (1997). 

\bibitem{chemin2017} 
Ars\`ene Chemin, Pauline Besserve, Aude Caussarieu, Nicolas Taberlet, and Nicolas Plihon, 
``Magnetic cannon: The physics of the Gauss rifle,''
Am. J. Phys. \textbf{85}, 495 (2017).
%http://doi.org/10.1119/1.4979653 

\bibitem{unimainz2017} 
\protect\url{http://www.prisma.uni-mainz.de/deu/2270.php}.

\bibitem{nagrial1993} 
Mahmood H.\ Nagrial,
``Design Optimization of Magnet Couplings Using High Energy Magnets,''
Electr. Machines and Power Systems \textbf{21}, 115--126 (1993).
%http://www.tandfonline.com/doi/citedby/10.1080/07313569308909638 

\bibitem{schonke2015b} 
Johannes Sch\"onke, 
``Smooth Teeth: Why Multipoles Are Perfect Gears,'' 
Phys. Rev. Appl. \textbf{4}, 064007-1--064007-9 (2015).  

\bibitem{jugendforscht2017} 
\protect\url{http://www.ignaz-guenther-gymnasium.de/downloads/scichallenge_1.pdf}.

\bibitem{jackson} 
John David Jackson, 
\textit{Classical Electrodynamics}, 3rd edition (Wiley, New York, 1998).

\bibitem{volkel2016} 
Simeon V\"olkel,
\textit{Rastmomentfreie kontaktlose Kupplungen aus kugelf\"ormigen Permanentmagneten}, 
Master thesis, University of Bayreuth (2016).

\bibitem{borgers2016} 
Stefan F.\ G.\ Borgers,
\textit{Bestimmung des Rastmoments kontaktloser Kupplungen aus kugelf\"ormigen Permanentmagneten}, 
Bachelor thesis, University of Bayreuth (2016).

\bibitem{schonke2016} 
Johannes Sch\"onke, Wolfgang Sch\"opf, and Ingo Rehberg,
``Magnetkugeln --- ein 10-Euro-Labor,'' 
Physik Journal 15 \textbf{4}, 31--37 (2016).  

\end{thebibliography}
\end{document}